\documentclass{iau}

\usepackage{amsmath}
\usepackage{graphicx}
\usepackage{multirow}

\begin{document}

\lefttitle{Yusef-Zadeh}
\righttitle{Tests of Sgr A*'s State}

\jnlPage{1}{7}
\jnlDoiYr{2021}
\doival{10.1017/xxxxx}

\aopheadtitle{Proceedings IAU Symposium}
\editors{M. Zaja\v{c}ek,  T. Je\v{r}\'{a}bkov\'{a}, V. Karas, R. Schödel \&  P. Sukov\'{a}, eds.}

\title{Observational Tests of Sgr A*'s\\
State of the Accretion Flow}

\author{F. Yusef-Zadeh$^{1,2}$, R. G. Arendt$^{3,4}$, H. Bushouse$^5$, C.~J. Chandler$^6$, D. Kunneriath$^7$, S. M. Ressler$^8$, 
M. Wardle$^9$}
\affiliation{$^1$Dept Physics and Astronomy, CIERA, Northwestern University, 2145 Sheridan Road, Evanston,
IL 60207, USA}
\affiliation{$^2$Observatoire de Paris, LUX, CNRS, PSL University, Sorbonne University, 75014, Paris, France}
\affiliation{$^3$Code 665, NASA/GSFC, 8800 Greenbelt Road, Greenbelt, MD 20771, USA}
\affiliation{$^4$UMBC/CRESST 2}
\affiliation{$^5$Space Telescope Science Institute, 3700 San Martin Drive, Baltimore, MD 21218}
\affiliation{$^6$NRAO, P.O. Box O,  Socorro, NM 87801, USA}
\affiliation{$^7$NRAO, 520 Edgemont Road, Charlottesville, VA 22903, USA}
\affiliation{$^8$ CITA, University of Toronto, Toronto, On, Canada MSS 3H8}
\affiliation{$^9$School of Mathematical and Physical Sciences, Astrophysics and Space Technologies Research Center,
Macquarie University, Sydney NSW 2109, Australia}
\begin{abstract}
General relativistic magneto-hydrodynamic (GRMHD) 
simulations of the possibly magnetically arrested disk (MAD) of the accretion flow of Sgr A*  predict short and long time 
variability, lumpiness in the flow, strong magnetic fields,  reconnection driving  flux eruptions and a jet 
aligned with  its  spin angular momentum. 
We carried out 
simultaneous multi-wavelength observations with the objective of placing some constraints on these predictions. Our  
analysis supports most of the predictions of the  MAD picture of the flow.  
It is not clear if there is evidence for a jet,  but this could be the result of 
confusion in the complex region surrounding Sgr A*. 
Also, the geometry of the magnetic fields, which accelerate particles to 
produce X-ray emission via inverse Compton Scattering (ICS), 
appears to be  consistent with a hybrid state  where  the accretion flow is represented 
 by MAD-like magnetic field strength and with 
Standard and Normal Evolution-like (SANE-like)  magnetic field geometry. 
\end{abstract}

\begin{keywords}
Black hole, Accretion disk, Magnetic Field, Acceleration
\end{keywords}

\maketitle

\section{Introduction}
Sgr A*, due to its relatively small mass ($\sim 4\times 10^6 {\rm M}_\odot$)
 compared to other supermassive black holes, offers a distinct observational 
advantage in that the physical processes around it occur on much shorter timescales. The orbital timescale close to the event horizon is 
only a {\it minute} instead of $\sim10^3$ min for more massive AGN black holes. All the time scales are reduced, making it feasible to use 
temporal variations to measure physical parameters close to the event horizon and compare them with numerical simulations.

GRMHD simulations of the accretion flow attempting to explain the origin of the variability of Sgr A*, which is localized in the innermost 
region of the accretion flow \citep{gravity18}, fall into two broad classes defined by magnetization and inflow dynamics. Magnetically Arrested 
Disk (MAD) models predict strong magnetic fields which suppress small-scale turbulence but trigger multiple stochastic eruptions per day, which 
are driven by magnetic reconnection that  accelerates particles to high energies
 \citep{dexter20,ripperda20,ressler23,aimar23,jiang24,dimitro25}. 
 A  collimated jet is also expected for a  black hole with non-zero spin. 
In MAD
models, the accretion disk brings in a large amount of coherent magnetic flux and the magnetic pressure becomes so strong that
it pushes back against the infalling gas, arresting the accretion flow. In this picture, the magnetic field lines are anchored to
the event horizon and twisted by the black hole's rotation. This creates an electromagnetic funnel that launches a relativistic jet,
provided that the black hole has a non-zero spin ($a > 0$ where $a$ is a dimensionless parameter related to the angular momentum of the black hole).
On the other hand, in 
SANE models, turbulence and short time scale variability dominate.   
The magnetic fields are relatively weak and the gas falls
into the black hole without much resistance from magnetic pressure, thus a jet is typically not produced or is extremely weak \citep{narayan12}. 
The magnetic fields are relatively weak and the gas falls
into the black hole without much resistance from magnetic pressure, thus a jet is typically not produced or is extremely weak \citep{narayan12}.
It is important to note that the MAD and SANE accretion-flow models represent two extreme cases. 
The accretion flow in Sgr A* may instead correspond to an intermediate or hybrid state between these two extremes, as discussed in \S5.

The difference between MAD and SANE flows imprints itself in four aspects that can be tested observationally:
{\it (a)} 
The presence of dramatic recurrent eruptions in MAD disks at a rate  of a few times a day that can  generate large-scale lumpiness or disturbance to the accretion disk \citep{murchikova22,ressler23}. 
{\it (b)} MAD disks display less short timescale turbulence than 
SANE disks  due to the suppression of the magnetorotational instability (MRI) in MAD; 
{\it (c)} in MAD disks strong magnetic flux 
builds up linearly until a critical saturation point is reached where magnetic reconnection in the mid-plane and jet sheath region 
of the accretion disk accelerate 
particles to high energies; {\it (d)} a strong collimated jet with outflow energies greater than $\dot{M} c^2$
is expected if the black hole has a non-zero spin.

To place some constraints on  the predictions made from GRMHD simulations, we carried out multi-wavelength observations to monitor the variability of Sgr 
A*. Seven epochs of JWST observations provided a gold mine to characterize the variability of Sgr A* in 2023 and 2024 \citep{zadeh25,zadeh26}. 
Here we use the 
results of these measurements to investigate the properties of the accretion flow of Sgr A*, 
as predicted from GRMHD simulations. Such simulations have already had considerable success in explaining certain observational features of Sgr A*
 \citep{ripperda20,dexter20,ressler23}.

\section{Power spectrum of the flux variability}

The power spectra of Sgr A*  
in Figure 1a taken at one particular epoch
 show two distinct regimes characterizing short and long time scales.  
The power spectra of the light curves of Sgr A* indicate non-stop flaring activity on minute to hourly time scales, 
which probe event-horizon length-scales (see Fig. 1a). 
Using the seven epochs of JWST observations, we obtained multiple power spectra that display slopes ranging from $-2.70$ to $-3.57$.
The steepness of the slope of the power spectrum  is indicative of the degree to 
 which the accretion disk is non-uniform and lumpy. 
Figure 1a shows a power spectrum slope of $\sim-$2.8. Detailed simulations of MAD flow indicate that the accretion disk of Sgr A* is quite lumpy 
whenever there is an outburst of magnetic flux. The evidence  for varying slope of the power spectrum 
is  consistent with the MAD picture in which flux eruptions are generated randomly. 
This is consistent with a lack  of periodicities (i.e. no spikes in power at any specific  frequencies) in all epochs of observations.
SANE accretion flows are   expected to be smooth. 

  \begin{figure}
\centering
    \includegraphics[scale=0.2]{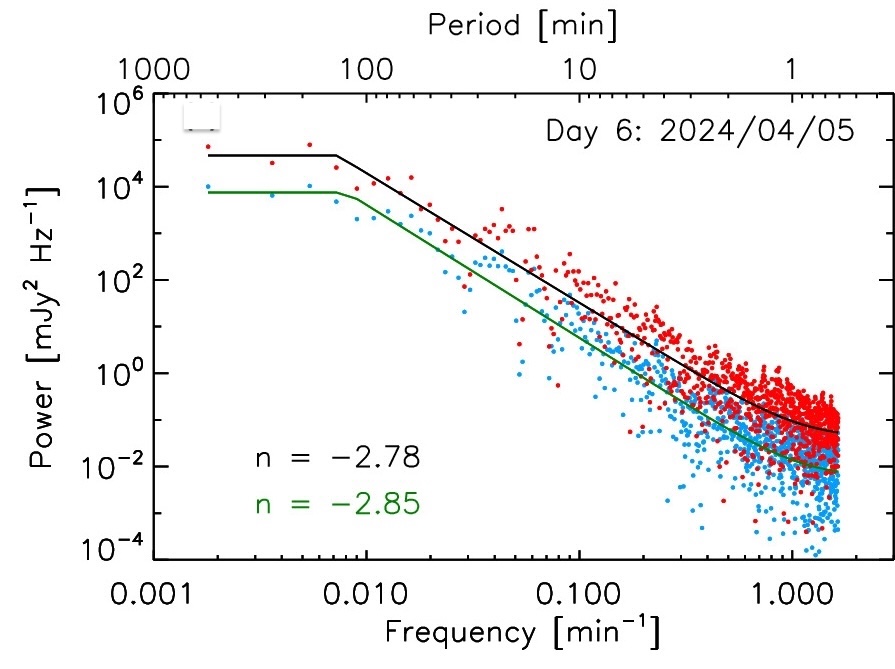}
\hspace{0.2in}
    \includegraphics[scale=.37]{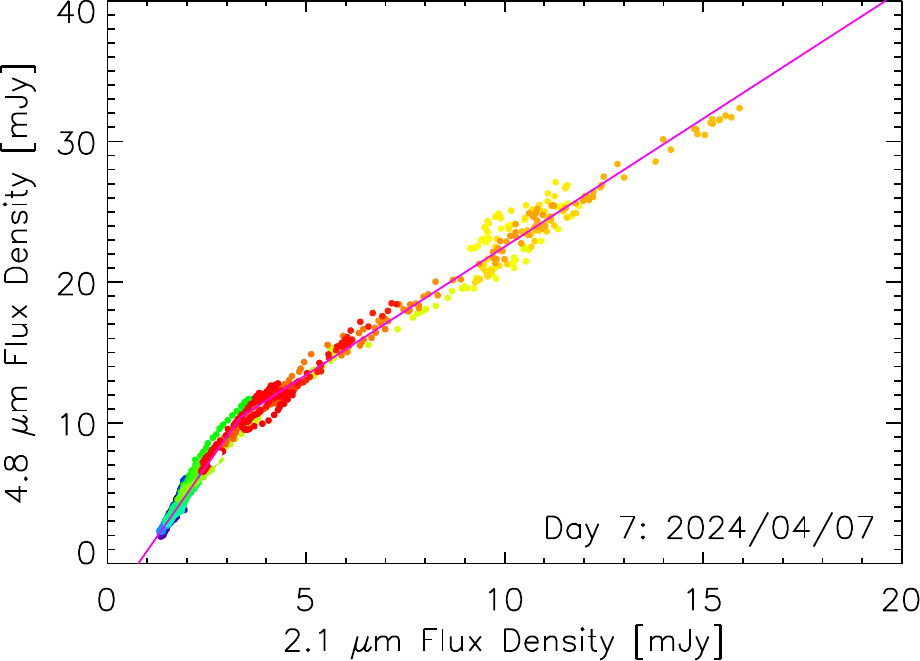}\\
    \vspace{0.2in}
\includegraphics[scale=.25]{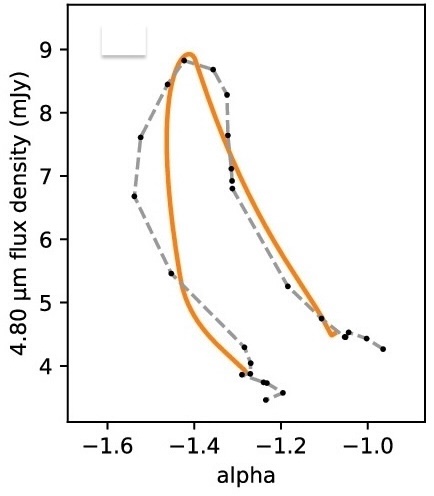}
\hspace{0.2in}
    \includegraphics[scale=0.4]{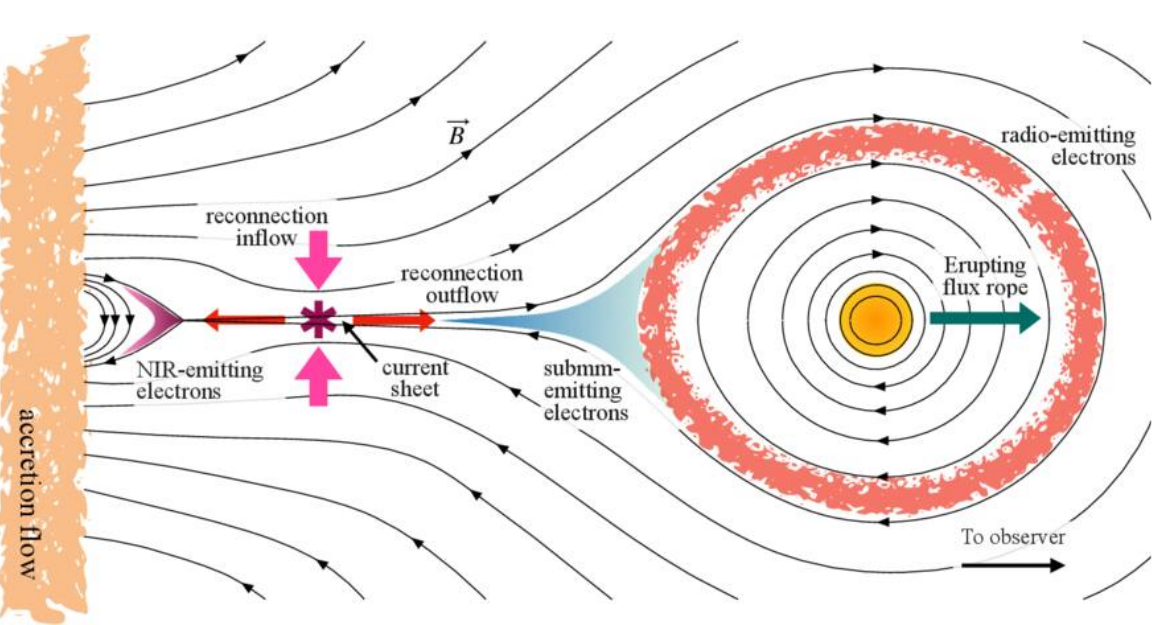}
    \caption{
{\it (a) Top Left}
Power spectrum  of the temporal variations of Sgr A* at one particular epoch of observations.
The  data are fitted with broken power-law trends with a transition
frequency below which  the power-law distribution becomes flat.
The slope of the power law $n$ with  its red noise characteristic 
varies  daily,   highlighting the random nature of flaring  activity
in the accretion flow. 
{\it (b) Top Right}
A scatter plot of  flux densities 
 at 4.8$\mu$m against those at 2.1 $\mu$m 
indicating a break where  the spectral index changes. 
 The weaker IR emission  have steeper spectral index 
 whereas the stronger  IR emission  show shallower spectrum. 
The change in the slope
 suggests two different populations with different spectral indices.
{\it (c) Bottom Left}
The variations of the spectral index as a function of brightness is represented as 
a  loop diagram which is modeled to be  due to synchrotron cooling, as the 4.8 $\mu$m emission lags in time behind the
2.1 $\mu$m emission.
The fit provides an estimate  of the strength of the  magnetic field, in this case 92.7 G.
{\it (d) Bottom Right}
A Schematic diagram showing production of X-ray emission by  ICS of NIR photons.
Magnetic reconnection  along  a flux rope accelerates particles as they get pushed out from the reconnection point. 
Ejected particles toward the accretion disk of Sgr A* (orange)
produce  enhanced NIR synchrotron emission, which is upscattered to X-rays by the particles in the accretion disk \citep{zadeh25,zadeh26}.
}
  \end{figure}

\section{Low-level fluctuations in the light curves}
Sgr A* shows 
both bright flares on long time scales and low-level fluctuations on short-time scales. 
A remarkable characteristic  that distinguishes low-level non-stop fluctuations from bright flares is their 
spectral index (see Fig. 1b).  
The unique capability of NIRCam to  observe two NIR wavelengths simultaneously,
allows the spectral index $\alpha$, where 
$\alpha \propto \log\, {F_\nu / \log\,{\nu}}$,
of bright and faint flares to  be determined as a function of time.
The strongest flaring activity in Fig. 1b tends to
show a shallower slope than the low-level fluctuations. 
This  striking result indicates that the spectral index 
becomes  shallower as the flux density increases.  
It is not clear what the origin of the low-level fluctuations is.  
Past simulations have considered that the  
short time-scale variability is consistent  with either  the MAD  or SANE torus models \citep{murchikova22}.

It is also possible that the population of low-level fluctuations at IR wavelengths is 
due to turbulent fluctuations which may be consistent with the SANE model,  whereas 
brighter populations of flux with a shallower spectral indices  may represent flux eruptions in the MAD picture.
There are limited  simulations of near-IR variability of Sgr A* \citep{grigorian24},
so our analysis  
can not distinguish between any of the models stated here.

\section{Magnetic fields of flares}
One of the advantages of simultaneously observing Sgr A* at 2.1 and 4.8 $\mu$m with JWST is  the ability to  
cross correlate [g] the light curves; doing so indicates a  time delay of 3 - 40 sec. 
This phase shift leads
to loops in plots of flux density vs spectral index as the emission rises and falls (see Fig. 1c). 

Using our observations, we measured the flux density as a function of the spectral index for both rising and falling stages of flares.
The variation of the spectral index trace counterclockwise loops (see Fig. 1c).
We explain the loop diagram using synchrotron
cooling of a population of electrons with a  power law energy distribution. 
Fitting this model to the data, we find 
that  
synchrotron cooling from a population of electrons  can explain  the shape of the observed
light curves and provide a direct estimate of the magnetic field strengths responsible for each flare. 
We find that for the clearest examples of flares, our model predicts that the magnetic fields have magnitudes in the range 40-90 G.
These values are  consistent with a MAD picture of the flow. 

\section{Acceleration mechanism: magnetic reconnection}

We carried out simultaneous JWST, NuSTAR, and VLA observations of Sgr A* on Apr 05, 2024.
 We reported the detection of a strong X-ray 
flare coincident with a bright IR flare, and a brightening in radio about an hour later \citep{zadeh26}.  We concluded that the X-ray flare can best be 
explained by inverse Compton Scattering (ICS)
of IR flare radiation that is beamed towards the accretion disk.  This  dynamic 
scenario is analogous to a coronal mass ejection \citep{lin24,zadeh26} 
in which a magnetic flux rope is ejected and reconnection produces oppositely directed flows 
moving upwards towards the rope to produce radio emission and downwards towards the accretion flow to produce X-ray emission due to ICS (see 
Fig. 1d). 
IR radiation from the approaching energetic electrons is enhanced by beaming and up-scattered by thermal electrons in the 
accretion flow to produce the strong X-ray flare. 

The  magnetic field geometry described in the above scenario to explain the X-ray emission 
 is more naturally consistent with a SANE picture of the accretion flow.  Recent works have found such structures in simulations where the accretion disk and jet have 
opposite polarities, resulting in frequent reconnection at the disk-jet boundary \citep{nathanail22}.  
This reconnection tends to remove electromagnetic energy from the jet and limit it strength.  
Although SANE accretion flows tend to have weaker magnetic fields as discussed in the previous sections, there may be a possible hybrid state between a SANE and MAD flow where the field strength is high 
but it is contained in several loop structures instead of a single, coherent direction (as in MAD flows).  Investigating this possibility in simulations is of key importance to makes sense of 
our observational findings.

\section{Collimated jet}

The high levels of linear and circular polarization observed at millimeter wavelengths with the Event Horizon Telescope (EHT) 
near the ``shadow" of the black hole
indicate  ordered magnetic fields capable of resisting gravity and potentially powering a jet.
There have been additional  indirect claims for the presence of a  jet from Sgr A*,
 however, none has provided
direct evidence for a  jet originating from Sgr A*. This is mainly because the region surrounding Sgr A*
is densely populated with diffuse and
compact structures spanning radio to X‑ray wavelengths, rendering it confusion‑limited. One implication of the magnetic field geometry discussed 
in the previous section is the presence of current sheets driving an outflow along the jet sheath region. 
Clearly, additional detailed observations are needed to search for a jet arising from Sgr A* and to test the MAD picture of the accretion flow.

\end{document}